\documentstyle[12pt,epsf,epsfig]{article}  
\def\Z{{\cal Z}}
\def\li2{{\rm Li}_2}

\def\roughly#1{\,\,\raise.3ex\hbox{$#1$\kern-.75em\lower1ex\hbox{$\sim$}}\,\,}

\def\eightpisq{{1\over 8\pi^2}}

\def \lsim{\mathrel{\vcenter
     {\hbox{$<$}\nointerlineskip\hbox{$\sim$}}}}
\def \gsim{\mathrel{\vcenter
     {\hbox{$>$}\nointerlineskip\hbox{$\sim$}}}}

\def\fo{\hbox{{1}\kern-.25em\hbox{l}}}

\def\bea{\begin{eqnarray}}
\def\eea{\end{eqnarray}}
\def\beq{\begin{equation}}
\def\eeq{\end{equation}}
\def\eq{\end{equation}}
\def\to{\rightarrow}

\def\bsg{\ifmmode B\to X_s\gamma\else $B\to X_s\gamma$\fi}
\def\bsll{\ifmmode B\to X_s\ell^+\ell^-\else $B\to X_s\ell^+\ell^-$\fi}
\def\bstt{\ifmmode B\to X_s\tau^+\tau^-\else $B\to X_s\tau^+\tau^-$\fi}
\def\shat{\ifmmode \hat{s}\else $\hat{s}$\fi}

\newcommand{\newc}{\newcommand}

\newc{\lcal}{\int {\cal L}dt}
 
\newc{\LSP}{{\chi^0_1}}
\newc{\stauR}{{\tilde \tau_R}}
\newc{\stau}{{\tilde \tau_1}}
\newc{\mstop}{m_{\tilde{t}}}
\newc{\mHpm}{m_{H^\pm}}
\newc{\ie}{{\it i.e.}}          
\newc{\etal}{{\it et al.}}
\newc{\eg}{{\it e.g.}}          
\newc{\kev}{\hbox{\rm\,keV}}            
\newc{\mev}{\hbox{\rm\,MeV}}            
\newc{\gev}{\hbox{\rm\,GeV}}            
\newc{\tev}{\hbox{\rm\,TeV}}
\newc{\xpb}{\hbox{\rm\, pb}}
\newc{\xfb}{\hbox{\rm\, fb}}

\newc{\mtop}{m_t}
\newc{\mbot}{m_b}
\newc{\mz}{m_Z}
\newc{\mw}{M_W}
\newc{\alphasmz}{\alpha_s(m_Z^2)}
\newc{\swsq}{\sin^2\theta_W}
\newc{\tw}{\tan\theta_W}
\newc{\cw}{\cos\theta_W}
\newc{\sw}{\sin\theta_W}
\newc{\BR}{\hbox{\rm BR}}
\newc{\zbb}{Z\to b\bar}
\newc{\Gb}{\Gamma (Z\to b\bar b)}
\newc{\Gh}{\Gamma (Z\to \hbox{\rm hadrons})}
\newc{\rbsm}{R_b^\hbox{\rm sm}}
\newc{\rbsusy}{R_b^\hbox{\rm susy}}
\newc{\drb}{\delta R_b}

\newc{\sgn}{\mbox{sgn}}
\newc{\tbeta}{\tan\beta}
\newc{\uL}{{\tilde u_L}}
\newc{\uR}{{\tilde u_R}}
\newc{\cL}{{\tilde c_L}}
\newc{\cR}{{\tilde c_R}}
\newc{\tL}{{\tilde t_L}}
\newc{\tR}{{\tilde t_R}}
\newc{\dL}{{\tilde d_L}}
\newc{\dR}{{\tilde d_R}}
\newc{\sL}{{\tilde s_L}}
\newc{\sR}{{\tilde s_R}}
\newc{\bL}{{\tilde b_L}}
\newc{\bR}{{\tilde b_R}}
\newc{\eL}{{\tilde e_L}}
\newc{\eR}{{\tilde e_R}}
\newc{\mhp}{m_{H^\pm}}
\newc{\mhalf}{m_{1/2}}
\newc{\emt}{{e/\mu /\tau}}

\def\lappeq{\mathrel{\rlap{\raise.5ex\hbox{$<$}}
{\lower.5ex\hbox{$\sim$}}}}
\def\gappeq{\mathrel{\rlap {\raise.5ex\hbox{$>$}}
{\lower.5ex\hbox{$\sim$}}}}

\newcommand{\drawsquare}[2]{\hbox{%
\rule{#2pt}{#1pt}\hskip-#2pt
\rule{#1pt}{#2pt}\hskip-#1pt
\rule[#1pt]{#1pt}{#2pt}}\rule[#1pt]{#2pt}{#2pt}\hskip-#2pt
\rule{#2pt}{#1pt}}

\newcommand{\Dal}{\drawsquare{7}{0.6}}

\begin{document}

\baselineskip=18pt

\begin{titlepage}
\begin{flushright}
CERN-TH/99-73\\
SNS-PH/99-03
\end{flushright}

\begin{center}
\vspace{1cm}

{\Large \bf 

Sparticle Masses from the Superconformal Anomaly}

\vspace{1.0cm}

{\large Alex Pomarol$^{1,}$\footnote{On leave of absence from
IFAE, Universitat Aut{\`o}noma de Barcelona, 
E-08193 Bellaterra, Barcelona.}
and Riccardo Rattazzi$^2$ }

\vspace{.5cm}

{\it $^1$ Theory Division, CERN\\
     CH-1211 Geneva 23, Switzerland}

\vspace{.3cm}
{\it $^2$ INFN and Scuola Normale Superiore\\
I-56100 Pisa, Italy}
\end{center}
\vspace{1cm}

\begin{abstract}
\medskip
We discuss a recently proposed scenario where the sparticle masses
are purely mediated by gravity 
through the superconformal anomaly. 
This scenario elegantly evades the supersymmetric
flavor problem since soft masses, like the anomaly, are
not directly sensitive to ultraviolet physics. However,
its minimal incarnation fails by predicting tachyonic sleptons. 
We study the conditions for decoupling of heavy
threshold effects and how these conditions are evaded. 
We use these results  to build a realistic
class of models where the non-decoupling effects of ultra-heavy vectorlike
matter fields eliminate the tachyons. 
These models have a flavor invariant superspectrum similar to that of
gauge mediated models. They, however,  differ in several aspects: 
the gaugino masses are not unified,  the colored sparticles are not 
much heavier than the others, the $\mu$ problem is less severe and
the gravitino mass is well above the weak scale, $m_{3/2}\gsim 10$ TeV.
We also show that in models where an $R$-symmetry can be gauged,
the associated $D$-term gives rise
to soft terms that are similarly insensitive to  the ultraviolet.

\end{abstract}

\bigskip
\bigskip

\begin{flushleft}
CERN-TH/99-73\\
March 1999
\end{flushleft}

\end{titlepage}

\section{Introduction}

Supersymmetry is at the moment the best candidate for physics
beyond the standard model (SM). It gives
a natural explanation of the gauge hierarchy and its simplest 
realization, the minimal supersymmetric standard model (MSSM),
remarkably satisfies the unification of the gauge forces.
However, superparticles have not been detected yet, 
nor any  deviation from the SM has been seen in precision
experiments and flavor physics. This means that not only supersymmetry
must be broken to give sparticles a mass, but also that
the resulting spectrum will most likely display a clever flavor structure.
The origin of supersymmetry breaking is thus crucial for phenomenology.

The simplest way to give superparticles a mass is certainly
via non-renormalizable interactions in supergravity \cite{gravmed},
as the basic ingredient, gravity, is a fact of life. 
Moreover gravity
mediation is generic as its agent (or the underlying more fundamental 
interaction) couples to everything: whatever dynamics breaks supersymmetry,
gravity will  eventually let us know. The flipped side of the coin is that
generic soft terms are not enough. They must also
be very specific to avoid large flavor violations. 
It is possible that this  problem
will  only be fully understood in a theory of quantum gravity
such as string theory. 
A  possibility though is that
some underlying flavor symmetry solves the problem by aligning 
the soft terms to the fermion masses \cite{flavsymm}. 

Models with gauge mediated supersymmetry breaking (GMSB) \cite{DNS,grrev}, 
on the other hand, offer a calculable and flavor invariant superspectrum.
These two properties motivated the recent activity on this scenario.

In this paper we focus on a 
particular
supergravity scenario  \cite{rs} in which the soft masses are
generated just by the auxiliary field of the gravitational multiplet.
This differs from previous ones \cite{gravmed} since no 
direct contribution arises from the hidden sector.
In this sense, soft masses are {\it purely} mediated by gravity.
This possibility was advocated in ref.~\cite{rs} in theories
with extra dimensions where only gravity propagates. Moreover the
same results, but limited to gaugino masses and $A$-terms,
are inevitably obtained in any
model where the hidden sector breaks supersymmetry dynamically in
the absence of singlets \cite{glmr}. 
From the supergravity
point of view, this idea corresponds to a no-scale
form for the Kahler potential \cite{noscale}
and to  canonical tree-level gauge kinetic
terms. Both scalar and gaugino masses are
zero at tree level and are purely determined by quantum effects.
It is easy to show that these quantum effects are just dictated
by the (super)conformal anomaly \cite{rs,glmr}, so that we 
can rightfully
name this scenario anomaly mediated supersymmetry breaking (AMSB).
As anomalies only depend on the low-energy effective theory,
so will soft terms. 
Soft masses at a scale $\mu$ will be written
as functions of the couplings at $\mu$, 
with no additional ultraviolet
(UV) dependence.
This 
property is very interesting for the supersymmetric
flavor problem.  In general soft masses
can pick up all sort of dangerous contributions 
from intermediate thresholds,
as they flow to the infrared (IR).
AMSB, however,  provides special trajectories
where all these potentially 
dangerous effects 
manage to disappear at low energy.
Unfortunately the AMSB scenario cannot be applied to the MSSM,
as it leads  to tachyonic sleptons \cite{rs}.  
The goal of the present paper is to construct realistic models 
(without tachyons) where
the original source of soft terms is the conformal anomaly.
Our basic point is that an intermediate threshold 
can displace  soft terms  from
 the AMSB 
trajectory, provided that the threshold position
is controlled by a light field (modulus). Using this remark
we will build models where the intermediate threshold is provided by
a messenger sector similar to that in  GMSB model. The interesting
thing is that, unlike GMSB,
the present mechanism works even in the absence
of tree-level mass splitting inside the messenger supermultiplets!
For this reason we will call this mechanism {\it anti}-gauge-mediation.

This paper is organized as follows. In section 2 we review AMSB,
relying mostly on a 1PI definition of the running soft terms. Section
3 focuses on decoupling and non-decoupling of heavy threshold 
and the problems with tachyon states.
Section 4 and 5 are devoted
to the construction of explicit models and to
a brief discussion of their phenomenology, 
while section 6 provides 
examples of solutions to the $\mu$-problem. Section 7 is somewhat outside
the main line of the paper. There we show that there is a more general
family of non-trivial RG trajectories which is UV insensitive.
It is interesting that this  class can be determined
by internal consistency in the flat limit 
or simply by gauging
an $R$-symmetry in supergravity.
Finally section 8 contains our conclusions.

\section{Anomaly mediated supersymmetry breaking}

The crucial feature of the scenario we  consider is that
supersymmetry breaking effects in the observable sector have
a pure gravitational origin. This is equivalent to saying that
the only relevant source of soft terms 
is the auxiliary field $u$ of the off-shell
gravitational supermultiplet for Poincar\`e supergravity. The 
natural formalism to describe these effects is the superconformal
calculus formulation of supergravity \cite{sugra}. 
In this framework one introduces
a chiral superfield $\phi$ with Weyl weight $\lambda=1$ playing
the role of the compensating multiplet for super-Weyl transformations.
With the aid of $\phi$ it is relatively easy to write a locally 
superconformal invariant lagrangian. Poincar\`e supergravity is then recovered
by fixing the extraneous degrees of freedom by a suitable set of gauge
conditions. In particular one can make the auxiliary field $u$ to reside only
in $\phi$ by fixing $\phi=1+\theta^2u/3$. (We will later comment on
different choices of super-Weyl-Kahler gauge fixing). 
Then the effects of $\langle u\rangle/3=F_\phi\not = 0$ on any operator are simply 
determined by inserting the suitable powers of $\phi, \phi^\dagger$ 
that render the operator Weyl invariant. It is convenient to assign 
to each physical field a Weyl weight equal to its mass dimension. Then the most
general Weyl invariant action has the form
\beq
\int d^4\theta \phi\phi^\dagger {\cal K}\left({\phi^{1/2}\over \phi^\dagger}
D_\alpha, {Q\over \phi},
{W_\alpha\over \phi^{3/2}},V\right ) +{\rm Re}\int d^2\theta \phi^3 
{\cal W}\left ({Q\over \phi},{W_\alpha\over \phi^{3/2}}\right )\, ,
\label{covar}
\eeq
where by $Q$ and $W_\alpha$ we collectively indicate the matter
and gauge field strength chiral superfields. For simplicity
we have omitted the dependence of ${\cal K}$ on $\bar D_{\dot \alpha}$,
$Q^\dagger$, etc..              

A few comments on eq.~(\ref{covar}) are in order.
We have omitted all terms involving covariant
derivatives acting on $\phi$, $\phi^\dagger$ (curvature terms). This is 
because we are only concerned with the RG evolution of soft terms.
By simple power counting, terms involving  $D_\alpha \phi$, $D^2\phi$, etc. 
cannot arise from ultraviolet divergences. This fact was already stressed
in ref.~\cite{aglr}. Indeed one could define the soft terms
at a scale $\mu$ by considering the superspace 
1PI action at external Euclidean momenta  
$p=\mu\gg m_{3/2}\sim F_\phi$ but truncated to terms with no
covariant derivatives acting on $\phi$. The neglected derivative
terms correspond
to the contribution from virtual momenta in the infrared domain
$m_{3/2}\lsim p\lsim \mu$ \footnote{Notice that in component notation
the  1PI is less useful to distinguish UV and IR contributions. For instance, 
in the softly broken Wess-Zumino model, the IR cut-off of the
tadpole diagram correction to the scalar mass is the soft 
mass itself, not $\mu$. 
In superspace language virtual momenta above and below $\mu$ contribute
to the coefficient of different operators.}.

One can easily check that the rule $D_\alpha\to  D_\alpha
\phi^{1/2}/\phi^\dagger$,
applied to $W_\alpha =\bar D^2D_\alpha V$ does give $W_\alpha \to W_\alpha/
\phi^{3/2}$. Moreover,
notice that the D'Alembertian is essentially $\Dal\sim D^2\bar D^2$, 
so that its 
Weyl covariant version is $\Dal/\phi\phi^\dagger$. This property
of $\Dal=-p^2$ crucially relates the soft terms to RG beta functions 
\cite{rs,glmr}. In order to elucidate this relation we recall
a few results.
As discussed in ref.~\cite{aglr}, in softly broken
supersymmetry, the soft terms associated to a chiral superfield $Q_i$
can be collected in a running superfield wave function $\Z_i(\mu)$ such that
\beq
\ln \Z_i(\mu)=\ln Z_i(\mu)+ \left (A_i(\mu)\theta^2+ {\rm h.c.}\right )
-m_i^2(\mu)\theta^2\bar\theta^2\, .
\label{defz}
\eeq
Here, $Z_i$ is the c-number wave
function,  $m_i^2$ is the sfermion mass, while $A_i$ contributes to $A$-terms
(for instance, a superpotential term $\lambda Q_1Q_2Q_3$ is associated
to $A_\lambda=A_1+A_2+A_3$). 
Now, consider the present theory in the
supersymmetric limit. 
The running wave functions can be defined
as $Z_i(\mu)=c_i(p^2=-\mu^2)$, where $c_i$ is the coefficient of
$Q_iQ_i^\dagger$ in the 1PI. 
Therefore, 
turning on $F_\phi$ simply amounts to
the shift $\mu^2\to \mu^2/\phi\phi^\dagger$
\beq
\Z_i(\mu)=Z_i\left({\mu\over {\sqrt {\phi\phi^\dagger}}}\right)\, .
\label{zshift}
\eeq
By using $\phi=1+F_\phi\theta^2$, we can write eq.~(\ref{zshift})
as
\beq
\ln \Z_i(\mu)=
\ln Z_i(\mu)-{\gamma_i(\mu)\over 2}\left (F_\phi\theta^2+
{\rm h.c.}\right )+
{\dot\gamma_i(\mu)\over 4} |F_\phi|^2\theta^2\bar\theta^2\, ,
\label{zsoft}
\eeq
where $\gamma_i$ is 
the anomalous dimension 
and $\dot\gamma_i=d\gamma_i/d\ln \mu$.
Comparing eq.~(\ref{zsoft}) to eq.~(\ref{defz}), we obtain
\begin{eqnarray}
A_i(\mu)&=&-\frac{\gamma_i(\mu)}{2}F_\phi\, ,\nonumber\\
m^2_i(\mu)&=&-\frac{\dot\gamma_i(\mu)}{4}|F_\phi|^2\, .
\label{softmasses}
\end{eqnarray}
We emphasize that this result 
is a simple consequence of the 1PI definition of the soft terms and of
the Weyl-covariantization property of $\Dal$.
This makes it clear that it does not depend on the regulator.
If we  had worked
in dimensional reduction (DRED) \cite{dred}, we would have gotten
the same result. 
This is because, in the bare
lagrangian, Weyl invariance requires the scale $\mu$ to always appear 
in the combination $\mu/{\sqrt {\phi\phi^\dagger}}$. Again,
 soft terms would be obtained as in eq.~(\ref{zshift}),
 by deforming the RG flow into superspace.

For gauge fields the relevant quantity 
is also a real superfield, $R(\mu)$, 
with components \cite{aglr}
\beq
R(\mu)= {1\over g^2(\mu)} -2{\rm Re}\left ({m_\lambda\over g^2}(\mu)\theta^2
\right ) +R_D\theta^2\bar \theta^2\, ,
\label{rfield}
\eeq
where $m_\lambda$ is the gaugino mass and where at lowest order in $g^2$ 
\beq
R_D=\eightpisq (T_G m_\lambda^2 -\sum _i T_i m_i^2)\, ,
\label{rd}
\eeq
where $T_G$ and $T_i$ are the Dynkin indices of the adjoint and matter 
irreducible representations. The real superfield $R(\mu)$
corresponds to the  physical 1PI coupling
and is related to the holomorphic coupling $S$ by \cite{nsvz,kl,ahm}
\beq
R=F\left(S+S^\dagger -\eightpisq\sum_i T_i\ln \Z_i\right )=F(\tilde R)\, ,
\label{RandS}
\eeq
where $F^{-1}$ can be expanded in a given scheme as
\beq
F^{-1}(x)= 1 -T_G\ln x/(8\pi^2)+\sum_{n>0} a_n/x^n\, .
\label{nsvz}
\eeq
The NSVZ scheme \cite {nsvz} corresponds to setting all $a_i=0$.

Similarly to the matter case, one starts from the kinetic coefficient
$1/g^2(\mu)$ in the supersymmetric limit and turns on $F_\phi$
by the shift $\mu^2\to \mu^2/\phi\phi^\dagger$:
\beq
R(\mu)= g^{-2}\left({\mu\over {\sqrt{\phi\phi^\dagger}}}\right )\, .
\label{rshift}
\eeq
Therefore, by using eq.~(\ref{rfield}), we have 
\beq
m_\lambda={g^2\over 2}{dg^{-2} \over d\ln \mu}F_\phi=
-{\beta(g^2)\over 2g^2}F_\phi\, .
\label{mlambda}
\eeq
It is also useful to consider the holomorphic coupling $S$, which, being
a chiral superfield, is defined by the shift
\beq
S(\mu)={1\over g_h^2({\mu\over \phi})}={1\over g_h^2(\mu)}-{b\over
8\pi^2}F_\phi \theta^2\, ,
\eeq
where $g_h$ is the holomorphic coupling in the supersymmetric limit
and $b$ is the 1-loop $\beta$-function coefficient.
Notice that although $S$ depends only on $\phi$, eq.~(\ref{RandS}) is
a function of the combination $\phi\phi^\dagger$, as it should be.
By using eqs.~(\ref{RandS}) and (\ref{nsvz}), it is easy to check that
our shift $\mu\to\mu/{\sqrt{\phi\phi^\dagger}}$, satisfies eq.~(\ref{rd})
at lowest order.
The same conclusion can be reached by using directly the gauge coupling
running at two loops in eq.~(\ref{rshift}). This is an important property
of $\mu\to\mu/{\sqrt{\phi\phi^\dagger}}$.
 Indeed, quite generally, and
regardless of supergravity, one may have asked whether a shift
$\mu\to\mu/{\sqrt{\phi\phi^\dagger}}$, applied to the running parameters of
a supersymmetric theory, does generate a RG trajectory for soft terms
({\it i.e.} whether the soft terms that are thus generated do solve their
RG equations, rather than being meaningless expressions). The property
we just mentioned is a crucial check, that our deformation does
generate a meaningful softly broken theory. We will see in sect. 7
that a similar argument can be used to accept or discard a more general
deformation of the supersymmetric RG flow.

Notice that, by eq.~(\ref{covar}), the soft terms associated to each operator
are essentially determined by the (quantum) dimension of its coefficient.
Indeed the soft masses of eqs.~(\ref{softmasses}) and
 (\ref{mlambda})  are just determined by the RG scaling
of the couplings, {\it i.e} by the conformal anomaly.
For a scale invariant theory, like for instance ${\cal N}=4$
Yang-Mills, the compensator dependence drops out in eq.~(\ref{covar}) and
 soft terms are not generated. 
This is why we refer to this scenario as 
``anomaly mediated supersymmetry breaking''.

We should recall that
in previous studies \cite{ibanez}, based on
the effective string action of Ref.~\cite{zwirner}, contributions
to gaugino masses proportional to the $\beta$-functions have been found.
However, in that case, unlike ours, the direct
source of the effect is the $F$-term of a modulus. Moreover, prior
to ref. \cite{rs}, no contribution to scalar squared-masses 
beyond one loop was discussed.

We now spend a few words on Kahler invariance versus
the physical meaning of $F_\phi$
\footnote{We thank Fabio Zwirner for raising and discussing this point.}. 
In the superconformal approach
the tree-level supergravity lagrangian is written as \cite{sugra,ku}
\bea
{\cal L}&=&-3\left [ e^{-K(Q,Q^\dagger)/3}\phi\phi^\dagger\right ]_D+
\left [ W(Q)\phi^3\right]_F=
\label{treesugra}
\\
&=&{1\over 2}\varphi\varphi^\dagger e^{-K(\tilde q,\tilde
q^\dagger)/3} R+{\dots}\, ,
\label{ricci}
\eea
where, for simplicity, we only display the Einstein term in the component
supergravity action. The two expressions in eq.~(\ref{treesugra})
reduce, 
in the absence of
gravitational fields,  respectively to $d^2\theta d^2 \bar\theta$
and $d^2 \theta$ integrals. 
Before fixing the
extraneous superconformal gauge freedom the compensator is a general 
chiral superfield
$\phi=\varphi+\chi\theta+F_\phi\theta^2$. Eq.~(\ref{treesugra}) is  
also invariant under Kahler transformations 
\bea
K&\to& K+f+\bar f\, ,\\
W&\to& e^{-f} W\, ,\\
\phi&\to&e^{f/3}\phi\, ,
\eea
where $f=f(Q)$ is a function of the chiral matter fields. As $\phi$ is not
 physical degree of freedom, Kahler invariance is not a true symmetry of
the lagrangian. Rather it states that physical quantities depend only on the
combination $G=e^KWW^\dagger$. Now, the quantity $F_\phi/\phi$,
on which we have been focusing so far, is not left invariant when $f$ involves
hidden sector fields with non-zero $F$-components. However also the
direct contribution from hidden fields to soft terms is not Kahler invariant,
but precisely compensates the change of $F_\phi/\phi$, leaving physical 
quantities unaffected. The scenario of pure mediation by $F_\phi$ that
we are considering then just corresponds to assuming the existence of a Kahler
``gauge'' where the hidden sector $F$-terms do not contribute to soft
masses in the observable sector. We may call this the ``convenient gauge''.
As the Kahler gauge is fixed, $F_\phi/\phi$
is now a physical quantity. As discussed in ref.~\cite{rs}, in the convenient 
gauge $K$ and $W$ split as
\bea
W&=&W(Q_h)+W(Q_o)\, ,\\
e^{-K/3}&=&1-f_h(Q_h,Q_h^\dagger)-f_o(Q_o,Q_o^\dagger)\, ,
\label{convenient}
\eea
where $Q_h$ and $Q_o$ denote respectively 
fields in the hidden and observable
sector. While the above form was motivated in ref.~\cite{rs} by a 
scenario with extra space-time dimensions, it just corresponds to the 
old no-scale ansatz for the Kahler potential \cite{noscale}.
It is interesting that this form of $K$ is stable under radiative
corrections due to non-gravitational interactions. A reflection of
this fact is that the parametrization for soft terms just in terms of
$F_\phi$ is valid at all renormalization scales. Again this has a beautiful
geometric explanation in the scenario of ref.~\cite{rs}. There the hidden and
observable sectors live on different ``branes'' separated by a bulk
where only gravity propagates: gauge and Yukawa interactions are relegated
to the observable brane and cannot induce mixings to hidden fields.

To conclude this little detour we give the expression of $F_\phi$
obtained by solving its equation of motion
\beq
{F_\phi\over \varphi}={1\over 3}K_iF^i+e^{K/3}{\varphi^{\dagger 2}\over 
\varphi}W^\dagger=
{1\over 3}K_iF^i+e^{K/2}W^\dagger\, .
\label{fphi}
\eeq
In the last equation we have made the Weyl gauge choice        
\beq
\varphi=\varphi^\dagger=e^{K(\tilde q,\tilde q^\dagger)/6}\, ,
\eeq
which renders the physics content of eq.~(\ref{treesugra})
more explicit, by making the Einstein term  canonical. Note that
the second contribution 
to $F_\phi$, $e^{K/2}W^\dagger$,  is just the gravitino
mass $m_{3/2}$, a Kahler invariant quantity. On the other hand,
the first contribution, $K_iF^i$, not only
should be evaluated in the convenient gauge but
is also model dependent. 
In the case that supersymmetry is broken
dynamically and without singlets in the hidden sector, 
we expect this first contribution to $F_\phi$
to be at least of ${\cal O}(m^{3/2}_{3/2}/M_P^{1/2})$, 
so that we can neglect it  \cite{glmr}.
In this interesting class of models the observable spectrum
is just fixed by the gravitino mass, and by gauge and Yukawa couplings.

\section{Decoupling and non-decoupling of high-energy thresholds
and the problem of tachyonic states}

In this section we will focus on the mass spectrum of AMSB models and
on the effects of intermediate mass thresholds.

As shown in the previous section our soft terms at a scale $\mu$ 
are determined by the
(super)conformal anomaly, which in turn is a property of the relevant
interactions of the effective theory at that scale. So it seems that,
by the way it was defined, in AMSB the UV thresholds affect
soft masses only indirectly, via their effects on gauge and Yukawa couplings.
Let us see this in more detail, by assuming that the theory  has a 
threshold at $M\gg F_\phi\sim m_{3/2}$. 
Let us also assume that there are no singlets below $M$.
Later we will see this is a crucial requirement.
An example satisfying these conditions would be the MSSM
with a vectorlike quark multiplet of mass $M\gg 1$ TeV. 
Now, it is easy to power count the $M$ dependence
of the terms in the low-energy effective action. 
Due to the absence of singlets, there can be no term
with a positive power $M$ (apart from a cosmological constant). 
Therefore the leading effects are logarithmic in $M$,
and affect the kinetic terms of the light fields at the quantum level
\footnote{
Higher-derivative operators scaled by $M^{-n}$, with $n>0$,
give a contribution to  soft terms suppressed 
by powers of $F_\phi/M\sim m_{3/2}/M$.
This class of effects can only be important when $M\sim m_{3/2}$.}.
Since $M$ is  a parameter in the lagrangian, it has Weyl weight
equal to zero. By Weyl invariance then the $M$ dependence of 
the wave functions,
eqs.~(\ref{zshift}) and (\ref{rshift}) ,
must take the form
\beq
\Z_i(\mu)=Z_i\left ({\mu\over{\sqrt{\phi\phi^\dagger}}}, 
M\right)\ ,\quad\quad
R(\mu)=g^{-2}\left ({\mu\over{\sqrt{\phi\phi^\dagger}}}, M\right)\, .
\label{decoup}
\eeq
This implies that 
the soft masses, that arise from the $\phi$-dependence of $\Z$ and $R$,
are still determined by the features of the low-energy theory
(at the scale $\mu$);
no extra contribution can come from the physics at $M$.

As an explicit example of this decoupling effect,
let us consider a sector with two fields $X$ and $Y$ and lagrangian
\beq
\int d^4\theta\, (XX^\dagger +YY^\dagger)+\int d^2\theta\, Y
(X^2-M^2\phi^2)\, ,
\label{XandY}
\eeq
where by convention we have taken $X$ and $Y$ with Weyl weight 1. 
The superfield $X$ vacuum expectation value is given by
\beq
\langle X\rangle=M\phi\label{xandphi}\, ,
\eeq
 up to irrelevant higher-derivative terms. Both 
$X$ and $Y$ become massive and can be integrated  out. 
After that, any low-energy wave function can depend 
on $\langle X\rangle$ but only through the
Weyl-invariant combination $\langle X\rangle/\phi=M$. Again
this does not affect the soft masses.
If we were to work with component fields, 
we would find that the threshold 
contributions from the heavy fields at $M$
combine themselves to preserve the AMSB form (eq.~(\ref{softmasses})) of the
low-energy soft terms.

It is rather clear why the AMSB scenario is interesting for the supersymmetric
flavor problem.  In the SM, 
the three Yukawa matrices are the {\it only}
relevant sources of flavor violation. This special property leads
to a natural suppression of dangerous flavor violating processes.
However this property is lost in a generic supersymmetric extension,
due to the presence of squark and slepton mass matrices that
are in principle  new sources of flavor violation. An elegant solution
to this problem is provided by models with GMSB.
In these models
the scale at which the three Yukawa matrices $Y_{u,d,e}$
are generated is well above the scale $\Lambda_S$ 
at which the soft terms are  induced. 
Therefore flavor violations in the sfermion masses 
are only  proportional to the low-energy $Y_{u,d,e}$
and we recover natural flavor conservation.
(Moreover the flavor invariant  contribution due to gauge interactions 
dominates these masses.)

The AMSB scenario also satisfies natural flavor conservation: scalar
masses are only functions of the low-energy
gauge couplings and  $Y_{u,d,e}$. 
Unlike gauge mediation, this property is rather independent
on where the scale of flavor is. 
As we explained above,
the soft masses in AMSB models decouple from UV physics,
and depend only on the IR theory. This
a great virtue, but unfortunately it is also what kills this scenario by
making it too (wrongly) predictive. 
The scalar masses are given by eq.~(\ref{softmasses}), that leads
in a  pure gauge theory to 
\beq
m^2_i(\mu)={c_ib\over 8\pi^2} \alpha^2(\mu) |F_\phi|^2\, , 
\label{neg}
\eeq
where $c_i>0$ is the quadratic Casimir of the scalar.
This contribution is therefore positive for asymptotically
free gauge theories ($b>0$) and negative for infrared free theories.
In the MSSM both SU(2)$_L$ and U(1)$_Y$ have $b<0$.
The sleptons, whose masses are essentially
determined by the SU(2)$_L\times$ U(1)$_Y$
gauge interactions
\footnote{Yukawa coupling effects are
negligible for (at least) the first two families.},  
are therefore tachyonic and the model is ruled out.

In ref.~\cite{rs},
 in order to avoid the tachyons, it was assumed
the presence of additional (non-anomaly-mediated) contributions 
to the scalar masses. These contributions
were associated to new fields propagating in the bulk. 
Although
these bulk contributions could lead to positive scalar masses, 
they  certainly
spoil the decoupling features of the AMSB models. 
These effects can be
associated to terms that mix  hidden with observable fields 
in eq.~(\ref{convenient}),
so that we apparently go back to the standard problem: in the absence
of an explicit model where the bulk effects are calculable
 the supersymmetric flavor problem remains open.

In the rest of the paper we will investigate the possibility of
 preserving the  property of natural flavor conservation in AMSB models, 
while avoiding  tachyons.
For this purpose,
we will reconsider the decoupling property of AMSB models.
As it is often the case, the absence of light singlets turns out to be 
necessary in the proof of decoupling.

Our basic point is the following. Suppose that we have a 
superfield $X$ such that the VEV $M$ of its scalar component 
sets the threshold.
If after the shift $X=M\phi+S$, the singlet $S$ is light, 
it must be kept in the effective theory.
In  this case the   Kahler potential can have a linear dependence on $M$:
\beq
\int d^4\theta\left (SS^\dagger +cM\phi S^\dagger+c^* M \phi^\dagger S
\right)\, ,
\label{tadpole}
\eeq
where $c$ is an ${\cal O}(1)$ number. 
In the presence of supersymmetry breaking these terms
are  important.
By minimizing we get 
a rather large $F_S=-c MF_\phi$. 
(Notice that this result corresponds
to the well known problem that singlets coupled to the MSSM Higgs
$H_1H_2$
can destabilize the weak scale because of
 their large $F$-terms \cite{ps}.) 
Thus  for a light singlet $S$, 
the proportionality of eq.~(\ref{xandphi})
 is violated (remember that 
$S$ has zero scalar VEV by assumption). 
The power 
counting is now drastically changed. 
For instance, an irrelevant operator
like
\beq
\int d^4\theta {SS^\dagger\over M^2} QQ^\dagger\, ,
\eeq
leads to an unsuppressed contribution $\sim |c F_\phi|^2$ to the mass
of the $Q$ scalar. 
Soft terms are no longer associated to the relevant
interactions of the low-energy theory and decoupling is lost.
This property of light singlets is the key ingredient of the
models we will construct. 
As we will see, in these new models, the tachyons can be eliminated by 
extra non-decoupling (flavor diagonal) contributions of heavy states.

Non-decoupling effects from light singlets 
have also been used in ref.~\cite{np} to lower
the effective scale of supersymmetry breaking.
The scenario in ref.~\cite{np} is different from ours
as it relies on a total singlet that   
 mixes at the Planck scale with the hidden sector fields.
Also, in that scenario $m_{3/2}$ must be much below the weak
scale, while in our case it is much above.

\section{Anti-gauge mediated supersymmetry breaking}

The first model 
we will present is  based on the following set of extra fields:
the  ``messenger'' fields, 
$\Psi$ and  $\bar \Psi$, in vectorlike representations
 of the SM gauge group, and the field
$X$, whose VEV  will give mass to the messengers.
The latter field has a very flat potential in order to
 fulfill the condition for non-decoupling that we  
explained above.
Like in gauge mediated models we take
$\Psi$ and $\bar \Psi$ 
to fit in complete SU(5) representations in order to preserve
gauge coupling unification. 
For instance we can take $N$ flavors of ${\bf 5}+
{\bf \bar 5}$.
The  superpotential of the model is given by
\beq
W=\lambda X \Psi\bar \Psi +{X^n\over {\Lambda^{n-3} \phi^{n-3}}}\, ,
\label{nonren}
\eeq
where again 
all fields have Weyl weight 1 and $n>3$.
The scale $\Lambda$ suppressing the non-renormalizable term
could conceivably be of the order of the Planck mass,
$\Lambda\sim M_P$. The form of the above superpotential could also
be enforced by an (anomalous) $R$-symmetry. 
Notice that the $\phi$-dependence of $W$ is dictated by
Weyl-invariance.
Now, when $F_\phi$
is turned on, the tree-level potential along the scalar-component
of $X$ becomes
\beq
V(X)=n^2\left |{X^{n-1}\over \Lambda^{n-3}}\right |^2+(n-3)
\left ({F_\phi X^n\over \Lambda^{n-3}}+{\rm h.c.}\right )\, .
\eeq
The minimum condition for $V(X) $ is 
\beq
{X^n\over |X|^2}=-{{n-3}\over {n(n-1)}}F_\phi^*\Lambda^{n-3}\, ,
\label{min}
\eeq
that leads to a non-zero VEV for X:
\beq
\langle X\rangle\sim m_{3/2}^{1\over n-2}\Lambda^{n-3\over n-2}\, .
\eeq
Eq.~(\ref{min})
also fixes the value of $F_X/X$ (the quantity relevant to soft terms)
\beq
{F_X\over X}= {nX^{*n}\over |X|^2 \Lambda^{*(n-3)}}={n-3\over n-1}F_\phi\, .
\label{fxox}
\eeq
Notice that, unlike the model of eq.~(\ref{XandY}),
 the VEV of the superfield $X$ is no 
longer proportional to $\phi$.
Since $X$ has weight 1, 
the low-energy wave functions depend on the threshold via 
$\tilde X\equiv X/\phi$, which has now a non zero $F$-component
\beq
{F_{\tilde X}\over \tilde X}=-{2\over {n-1}}F_\phi\, .
\label{ftild}
\eeq
Then in this model
the soft masses at low-energy are no longer
determined by the action of $d/d\ln \mu$ on the wave functions,
eqs.~(\ref{softmasses}) and (\ref{mlambda}), 
but by
\bea
{m_\lambda(\mu)\over g^2(\mu)}&=&{F_\phi\over 2}\left({
\partial\over \partial \ln\mu}+{2\over n-1}{\partial \over \partial\ln |X|}
\right ){1\over g^2(\mu,X)}\, ,\nonumber\\ 
A_i(\mu)&=&-{F_\phi\over 2}\left({
\partial\over \partial \ln\mu}+{2\over n-1}{\partial \over \partial\ln |X|}
\right )\ln Z_i(\mu,X)\, ,\nonumber\\
m_i^2(\mu)&=&-{|F_\phi|^2\over 4}\left({
\partial\over \partial \ln\mu}+{2\over n-1}{\partial \over \partial\ln |X|}
\right )^2\ln Z_i(\mu,X)\, .
\label{softmasses2}
\eea
From the component point of view, the soft terms in this model
are determined as follows. 
Down to the scale $X$ the soft terms respect the AMSB
form (\ref{softmasses}). 
At his scale, the messengers add a gauge mediated contribution.
This contribution does not adjust the soft terms to the AMSB trajectory
of the low-energy theory. For that to happen one would need
precisely $F_X/X=F_\phi$, which does not hold. Notice indeed
that decoupling is recovered only in the limit $n\to \infty$,
when the mass of the $X$ excitation around the minimum becomes very large.

As another example we now consider the previous model but in the absence
of the second term on the r.h.s. of eq.~(\ref{nonren}). 
In this case  
$X$ is a flat direction only lifted by the effects of $F_\phi\not =0$.
The relevant term along $X\not =0$ and $\Psi,\bar \Psi=0$ is
\beq
\int d^4\theta Z_X\left ({\sqrt{XX^\dagger/\phi\phi^\dagger}}\right ) 
XX^\dagger\, ,
\eeq
where $\mu^2\to XX^\dagger/\phi\phi^\dagger$ has been taken in the wave 
function,
since $X$ plays now the role of IR cut-off (notice that the above
equation satisfies Weyl symmetry).
At leading order the potential along $X$ is determined by the running soft mass
\beq
V(X)=m_X^2(|X|)|X|^2\simeq \left |{F_\phi\over 16\pi^2}\right |^2
N\lambda^2(X)\left [A\lambda^2(X)-C_ag_a^2(X)\right ] |X|^2\, ,
\label{runx}
\eeq
where $A,C_a>0$, and a sum over the gauge couplings $g_a$ of the messengers
is understood. 
The expression in square bracket is just the beta function
of $\lambda$. It is conceivable a situation where at a large energy scale
the $\lambda^4$ term dominates and $m_X^2>0$, while at lower values
of $X$ the gauge term becomes important leading to $m^2_X=0$ at
some $M$. The $X$ field is therefore stabilized at $\langle X\rangle
\sim M$.
This is just a supersymmetric version of the Coleman-Weinberg
mechanism \cite{cw}.
A situation like that is very easy to obtain if
the gauge group has an asymptotically free factor. In the MSSM only
SU(3) is weakly UV free and the presence of more than 3 flavors
of messengers makes it IR free. 
Nonetheless it is quite possible 
to gauge a subgroup of the messenger flavor group obtaining a 
strongly UV free factor. In this case the Coleman-Weinberg 
stabilization
of $X$ can work very well. 
$X$ could get a VEV
anywhere between the weak and GUT scales. 
Notice that from eq.~(\ref{runx}),
$F_X/X$ is a loop factor smaller that $F_\phi$,
 $F_X/X\simeq N\lambda^2F_\phi/16\pi^2$, and can be neglected.
Therefore  $F_{\tilde X}/\tilde X\simeq -F_\phi$ and the soft masses
are affected by the messenger threshold. This effect is given
by  eqs.~(\ref{fxox}) and (\ref{softmasses2})
extrapolated at $n=3$. 
The essential agreement between these two
limits is just due to the fact that they both correspond to classically
scale invariant models 
in which there are no soft terms at tree level.

\subsection{Superparticle spectrum}

A model like the one just outlined above is very similar to GMSB
models.
All sources of flavor violation that are active above the scale $X$
cannot affect the masses in a relevant way; this is because above
$X$ we are on the AMSB trajectory. 
Below the scale $X$, the soft terms are no longer on this 
privileged  trajectory and can pick up all sort of dangerous contributions.
Therefore we must assume that the scale of flavor $\Lambda_F$ (below 
which the only sources of mixings are $Y_{u,d,e}$) is somewhat larger 
than $X$. If the MSSM unifies in a simple gauge group, then $\Lambda_F$
cannot be bigger than the GUT scale $\sim 10^{16}$ GeV. 
Thus we will assume
from now on $X\lsim 10^{16}$ GeV.
For simplicity we will also consider the case in which the VEV of $X$
is fixed by the Coleman-Weinberg mechanism. 

In order to get the sparticle spectrum from eq.~(\ref{softmasses2}),
we only need to know the dependence of the gauge coupling $g$
and the wave function $Z$ on the scale $\mu$ and on the singlet $X$
induced after integrating out the 
messengers $\Psi$ and $\bar\Psi$ \cite{gr}. 
For a simple gauge group,
these are  given by 
\begin{equation}
\alpha^{-1}(\mu,X)=
\alpha^{-1}(\Lambda)+
\frac{b-N}{4\pi}\ln\frac{XX^\dagger}
{\Lambda^2}
+\frac{b}{4\pi}\ln\frac{\mu^2}
{XX^\dagger}\, ,
\label{gauge2}  
\end{equation}
and
\begin{equation}
Z_i(\mu,X)=Z_i(\Lambda)
\left(
\frac{\alpha(\Lambda)}{\alpha(X)
}\right)^{\frac{2c_i}{b-N}}
\left(\frac{\alpha(X)}{\alpha(\mu)
}\right)^{\frac{2c_i}{b}}\, .
\label{z}
\end{equation}
where $c_i$ is the quadratic Casimir. Using eqs.~(\ref{gauge2}),
(\ref{z}) and (\ref{softmasses2}) with $n=3$, we obtain
\begin{eqnarray}
m_\lambda(\mu)&=&\frac{\alpha(\mu)}{4\pi}
(b-N)
F_\phi\, ,
\label{gaugino}\\
A_i(\mu)&=&-\frac{2c_i}{4\pi}\left[\alpha(\mu)
+[\alpha(X)-\alpha(\mu)]\frac{N}{b}
\right]F_\phi\, ,\label{tri}\\
m^2_i(\mu)&=&
\frac{2c_ib}{(4\pi)^2}\left[\alpha^2(\mu)-\alpha^2(\mu)
\frac{N}{b}
+[\alpha^2(\mu)-\alpha^2(X)]\frac{N^2}{b^2}
\right]|F_\phi|^2\, .
\label{scalar}
\end{eqnarray}
At the scale $\mu=X$, we can see that 
the contributions (\ref{gaugino})-(\ref{scalar})  
arise from two sources. 
The first term is purely dictated by the superconformal
anomaly. The second one
is the effect of the   $N$ $\Psi$-$\bar\Psi$ threshold.
This second contribution 
is equal in magnitude but {\it opposite in sign} to the
usual in GMSB models. 
For this reason we call this scenario {\it anti}-GMSB.
Since it is well known that the GMSB contribution to the scalar
soft masses are positive, one could  think  that this will
not help us to solve the above problem of negative scalar masses. 
Nevertheless, this is not the case.
In {\it anti}-GMSB there is yet another contribution to
the scalar masses, the third term in brackets in 
eq.~(\ref{scalar}), originating from the gaugino mass via
RG.
Therefore in order to have positive scalar masses at $\mu\sim m_W$,
 the third term
of  eq.~(\ref{scalar}) must overcome the others.
The condition $m^2_i>0$ puts a lower bound on $N$ and $X$.
In figure~1 we plot the 
soft squared-masses 
of the left-handed squark (solid line),
 the right-handed down-squark (dashed-dotted line),
the left-handed slepton (dashed line) 
and the right-handed slepton (dotted line) normalized to 
 the SU(2)$_L$-gaugino 
squared-mass, as a function of Log$_{10} X$.
These masses are calculated at the scale $\approx$ 200 GeV.
We have taken $N=2,3,4$ pairs of  $\Psi$ and 
$\bar\Psi$ in the {\bf 5} and $\bf \bar 5$ representation
of SU(5).
We see that for $X\lappeq M_{GUT}\simeq 10^{16}$ GeV,
 one is forced to have $N\geq 3$ in order to have positive masses. 
The case $N=3$, however, 
implies a zero $\beta$-function for the SU(3) coupling
above $X$, 
and therefore zero gluino mass.
We then find that the only case that gives realistic soft masses corresponds 
to $N\geq 4$. As an example, let us consider the case $N=4$ with
$X=M_{GUT}$.
The spectroscopy of this scenario is completely different from anyone 
in the literature. 
Neglecting the effects of Yukawa couplings, we have
\begin{equation}
m_{\widetilde\lambda_1}\simeq 1.1\, m_{\widetilde\lambda_2}
\simeq 1.6\, m_{\widetilde\lambda_3}\simeq 1.6\, m_Q\simeq 
1.9\, m_L\simeq 
2\, m_U\simeq 
2.2\, m_E\simeq 
2.4\, m_D\, .
\label{masses}  
\end{equation}
We have not specified the 
 Higgsino mass since it depends on the $\mu$-parameter and therefore 
is model dependent. From eq.~(\ref{masses})
we see that the gaugino masses have an hierarchy opposite to that of
GMSB. 
Also notice that this scenario gives some of the squarks lighter than the
sleptons.
The stops and the Higgs $H_2$
receive an extra contribution proportional to the top 
Yukawa coupling $Y_t$ 
that is as important as eq.~(\ref{scalar}). 
For $H_2$  this is given by
\bea
\delta m_{H_2}^2(\mu)&=&{3\over (4\pi)^2}\alpha_t(\mu)\left\{\sum_i 
d_i\left [
-\alpha_i(\mu)+{2N\over b_i}(\alpha_i(\mu)-\alpha_i(X))+{F(\mu)\over E(\mu)}N
{\alpha_i^2(X)\over 2\pi}\right ]\right.\nonumber\\
&+&\left.{N^2\over 2\pi}{G_2(\mu)\over E(\mu)}+6\alpha_t(\mu)\left[1+
{N\over 2\pi}
{G_1(\mu)\over E(\mu)} \right ]^2\right \}{|F_\phi|^2}\, ,
\label{topcont}
\eea
where $i$ sums over the three MSSM gauge groups,
$b_i=(-6.6,-1,3)$, $d_i=(13/15,3,16/3)$,
$\alpha_t=Y_t^2/4\pi$ and  
\bea
E(\mu)&=&\Pi_i\left ({\alpha_i(t)\over \alpha_i(t_X)}\right )^{d_i/b_i}
\ ,\ \ t=\ln \mu\ ,\ \ t_X=\ln X\, ,\\
F(\mu)&=&\int_{t_X}^t E(t')d t'\, ,\\
G_1(\mu)&=&  \int_{t_X}^t E(t')\sum _i {d_i\over b_i}(\alpha_i(t')-
\alpha_i(t_X))
d t'\, ,\\
G_2(\mu)&=&\int_{t_X}^t E(t')\left \{\sum _i {d_i\over b_i}(\alpha_i^2(t')-
\alpha_i^2(t_X))+\left [\sum _i {d_i\over b_i}(\alpha_i(t')-\alpha_i(t_X))
\right ]^2\right \} d t'\, .
\eea
For 
the left-handed stop and right-handed stop, the top contribution
is obtained by just replacing the factor 3 in front of
eq.~(\ref{topcont}) by a factor 1 and 2 respectively.
We have checked that for values of the top coupling 
in the region 
$0.7\lappeq Y_t\lappeq 0.9$ 
the soft mass
$m^2_{H_2}$ is the only one negative. This can trigger 
electroweak symmetry breaking (EWSB).

The $\mu$-term  and the bilinear Higgs mass term,
$B\mu$, are not predicted by the model.
In the section 6, we will propose a way to generate them.
For phenomenological purposes, however,
they can be considered free parameters
of the theory. Their only constraint comes from EWSB.
We find  that EWSB can be  
achieved for moderate values of $\mu$, 
typically not bigger than the other soft masses. 
This is very convenient  since  it implies that 
the LSP in these theories can be  the neutral Higgsino. 
This is again different from
GMSB models where the  LSP is  the gravitino. Here the 
gravitino  gets a tree-level mass of order $F_\phi$ and becomes very heavy 
($\sim 10$ TeV).

\section{$D$-term contributions to scalar masses}

One could  also use the previous non-decoupling mechanism to
build  realistic models with extra U(1)'s at high-energies.
Extra U(1)'s are well motivated as they appear in GUT groups
of rank greater  than 5,
such as SO(10) or E$_6$. As we shall see,
additional contributions to the low-energy soft masses 
can be obtained from the $D$-terms of these U(1)'s.

Let us consider  two fields $\psi$ and $\bar\psi$
with U(1)-charges  1 and -1.
Let us also  assume that  the $D$-flat direction, 
$\psi=\bar\psi\equiv X$, is stabilized  by the Coleman-Weinberg
mechanism away from the origin, as in the previous section.
This will break the  U(1) giving a mass to the 
vector superfield proportional to $X$.  We can now 
integrate out the vector superfield. This induces, 
a new spurious dependence of the low-energy wave functions on $X/\phi$,
 and correspondingly a new contribution to the soft masses.
The mass of a scalar with U(1) charge $q_i$ will be corrected by
\footnote{There can be other contributions mediated by the $U(1)$
vector multiplet at the two-loop level \cite{gr}. 
These contributions, however, are proportional to $g^4_{U(1)}$
and can be in principle smaller than the ones here.}
\begin{equation}
\delta m^2_i=q_i\langle D\rangle
\, ,\ \quad\quad\ \ \ \langle D\rangle=\frac{1}{2}
[m^2_{\bar\psi}(X)-m^2_\psi(X)]\, ,
\label{dterm}
  \end{equation}
where 
$m^2_\psi$ and $m^2_{\bar\psi}$  are the anomaly mediated soft masses
 of $\psi$ and  $\bar\psi$, eq.~(\ref{softmasses})
\footnote{Notice that for $m^2_\psi=m^2_{\bar\psi}$ 
the $D$-term contribution 
eq.~(\ref{dterm}) is zero. Therefore $\psi$ and $\bar\psi$ 
must have different
couplings in order to generate the soft mass (\ref{dterm}).}.
This contribution to the slepton and squark masses
is   in general comparable  to that arising from the anomaly
eq.~(\ref{softmasses}). 
Its exact value depend on the RG trajectory of  
$Z_\psi$ and $Z_{\bar\psi}$ that is model dependent.
For a large enough value of $\langle D\rangle$
and $q_i>0$, the 
contribution (\ref{dterm}) can overcome the negative contribution
(\ref{softmasses}) and provide a realistic 
theory of soft masses.
The contribution (\ref{dterm}) is generated at the scale
at which the U(1) is broken.
This scale must be around the GUT scale if we do not
want to spoil the gauge coupling unification of the MSSM.
This does not necessary implies that the scale of
flavor must be well above the GUT scale to avoid
flavor violations in the soft masses.
As explained in ref.~\cite{dp},
if $q_i$ are flavor independent, the soft masses can  be flavor
diagonal at the leading order.

Gaugino masses are not affected by $D$-terms since the latter
are $R$-invariant.
Therefore in this scenario gaugino masses are predicted to have their
anomaly mediated value eq.~(\ref{mlambda}).

\section{The $\mu$-problem}

In order to have a   phenomenologically viable model,
a supersymmetric Higgs mass parameter $\mu$,  of the order 
of the soft terms, must be generated.
This is a well-known problem in GMSB theories where soft 
masses are induced by loop effects \cite{dgp}.
In our models, since soft masses are also induced at the one-loop level,
one might expect the $\mu$-problem to be equally severe.
On the contrary, we will show that in these theories
there is a simple way to generate the $\mu$ 
parameter. 
 
A priori, a  $\mu$ mass  could arise from the Weyl-invariant 
operator \cite{giumas}
\begin{equation}
\int d^4\theta \frac{\phi^\dagger}{\phi}H_1 H_2\, ,
\label{kahlermu}
\end{equation}
where $H_{1,2}$ are the two Higgs doublets of the MSSM.
This operator induces both $\mu$ and $B\mu$ at tree level.
While $\mu$ could be small because of a small coefficient 
multiplying eq.~(\ref{kahlermu}), this operator always
gives  $B=F_\phi$, which
is much larger than all the other soft masses. 
 Therefore this term must be forbidden.
The solution we propose is to generate the $\mu$-term by
the operator
\begin{equation}
\int d^4\theta\, H_1 H_2
 \frac{X^\dagger}{X}\tilde Z\left ({\sqrt{XX^\dagger/\phi\phi^\dagger}}
\right)\, , 
\label{muterm}
\end{equation}
where $\tilde Z$ is a wave function coefficient  at the 
renormalization scale 
$\mu_R\rightarrow {\sqrt{XX^\dagger/\phi\phi^\dagger}}$ 
(in this section we will denote the renormalization scale by $\mu_R$
instead of $\mu$  in order to distinguish it from the 
Higgs $\mu$-parameter),
and
$X$ is the light singlet with $F_X\simeq0$ introduced in section 4.

The operator (\ref{muterm})  generates a  one-loop $\mu$ term
and a two-loop $B\mu$ term:
\begin{eqnarray}
\mu&=&\left.\frac{1}{2}\frac{d \tilde Z}{d\ln \mu_R}
\right|_{\mu_R=|X|}\, ,\\
B\mu&=&\left.\frac{1}{4}\frac{d^2 \tilde Z}{(d\ln \mu_R)^2}\right|_{\mu_R=|X|}
\, .
\label{mubmu}
\end{eqnarray}
The operator (\ref{muterm})  can arise by integrating out a 
 singlet $S$  coupling to $H_1H_2$ and getting its mass from $X$.
A simple example is given by the superpotential
\begin{equation}
\int d^2\theta\left[ 
\lambda SH_1H_2+\frac{1}{3}kS^3+\frac{1}{2}y S^2X\right]\, .
\label{singlet}
\end{equation}
At one-loop, a kinetic mixing between $X$ and $S$ is
generated
\begin{equation}
\int d^4\theta 
\tilde Z(\mu_R)\, SX^\dagger+h.c.\,  .
\label{mixing}
\end{equation}
For a nonzero $X$ VEV, the singlet $S$ is massive and can be integrated out.
Its equation of motion is given by
\begin{equation}
S\simeq -\frac{\lambda}{y}\frac{H_1H_2}{X}\, ,
\end{equation}
which inserted in eq.~(\ref{mixing}) 
gives rise to eq.~(\ref{muterm}).

An alternative model is given by
\begin{equation}
\int d^2\theta
S(\lambda H_1H_2+\lambda_N N^2-\lambda_{\bar N}\bar N^2)\, .
\label{singlet2}
\end{equation}
This model has a flat direction along
$\lambda_N N^2=\lambda_{\bar N}\bar N^2\equiv X^2$.
Assuming as above that $X$ gets a VEV, 
the singlets get a mass.
Integrating them out, one obtains eq.~(\ref{muterm}) 
with  $\tilde Z\propto Z_{N}-Z_{\bar N}$.

We stress that in the present mechanism there is no danger
of generating a large $B$, which was instead the case for GMSB.
This is because nowhere in the visible sector
there is a supermultiplet with a tree-level mass splitting. 
All splittings
are of order $\alpha F_\phi\sim m_W$ to start with, 
and $B$ cannot come out bigger than that.

\section{Gauging an $R$-symmetry}

In the first section, independent of supergravity, we could have pointed out 
that the shift $\mu^2\to \mu^2/\phi\phi^\dagger$ 
is remarkable as it automatically
defines a deformation of the RG flow of a supersymmetric theory, into the
flow of a softly broken one. In this section we study under what conditions
one can define a more general deformation $\mu\to \mu e^{V/3}$, where $V$ is 
now a genuine vector superfield, and the factor 3 is chosen for later use. 
Indeed we will just need to discuss the case $V=V_D\theta^4$, since 
$\mu \to \mu \phi\phi^\dagger e^{V/3}$ parametrizes the most general shift.

Let us consider first a Wess-Zumino model with superfields $Q_i$, 
$i=1,\dots, N$,
and Yukawa couplings $\lambda_{ijk}$. To be slightly more general, we indeed 
define the soft terms by the formal shift
\beq
\Z_i(\mu)=e^{q_i V} Z_i\left (\mu e^{V/3}\right)\quad \longrightarrow
\quad m_i^2=-\left (q_i+{\gamma_i\over 3}\right )V_D\, .
\label{shiftwz}
\eeq
In order to check if these soft masses do represent an RG trajectory we
verify that they satisfy the evolution equations. By use of the results 
of ref.~\cite{aglr} the general RG equation at all orders has the form 
\beq
{d m_i^2\over d\ln \mu}=-{\partial \gamma_i\over \partial \ln \lambda_{klm}^2}
\left (m_k^2+m_l^2+m_m^2\right )\, ,
\label{rgwz}
\eeq
where summation over $k,l,m$ is understood. It is easy to check that
eq.~(\ref{shiftwz}) solves the one above if $q_k+q_l+q_m=0$ for any 
$\lambda_{klm}\not =0$. Notice that the $q$'s can thus be considered charges
of a background gauge symmetry, see eq.~(\ref{shiftwz}). Moreover, whatever the 
structure of
$\lambda_{klm}$, we can always define a consistent RG deformation with all $q$'s
vanishing. The situation changes when we turn to a gauge theory.
In addition to the shift of eq.~(\ref{shiftwz}), the superfield gauge couplings
are  defined by
\beq
R_a={1\over g_a^2\left (\mu e^{V/3}\right )}\, .
\label{gauge}
\eeq
We can do the same check we did for the WZ model. The Yukawa sector gives
the same constraint on the $q$'s as before. The constraint 
from the gauge sector is determined by eq.~(\ref{rd}) for $R_a|_D$
(remember that $R|_D$ is the coefficient of
 a non-local operator \cite{aglr}). By using eqs.~(\ref{shiftwz}) and 
(\ref{gauge}),
and $m_\lambda=0$, eq.~(\ref{rd}) implies
\beq
\left (b- \sum_i T_i\gamma_i\right ){V_D\over 3}=
\left (\sum_i T_i(-q_i-\gamma_i/3)
\right ){V_D\over 3}\, ,
\eeq
which, conveniently rewritten in terms of $r_i=q_i+2/3$, is 
\beq
N_c+\sum_i T_i(r_i-1)=0\, .
\eeq
This is just the condition for an $R$-symmetry of charges $r_i$
to be non anomalous. Notice also that the Yukawa constraint for
$r_i$ is also $r_i+r_j+r_k=2$ for each $\lambda_{ijk}\not = 0$. 
Thus we have established that the general deformation $\mu\to \mu e^V$
can only work for theories with a non-anomalous $R$-symmetry. Unfortunately
the MSSM does not have such a symmetry, so we cannot use this deformation
to improve the AMSB scenario. A possibility would be to add the suitable
matter multiplets at the weak scale \cite{gaugeR}. 

It is remarkable that by the above arguments in rigid supersymmetry
a gauged $R$ symmetry has emerged. This brings us back to our natural
arena: supergravity. Indeed, maybe less instructively,  we could have derived
this new deformation simply by gauging $R$ in supergravity. This is done in the
formalism of ref.~\cite{fgkv}, by introducing a connection $e^V$ under which
the compensator has charge $-2/3$, and matter fields have charge $r_i$.
The lagrangian is then given by
\beq
\int d^4\theta \phi\phi^\dagger 
e^{(r_i-2/3)V}Q_iQ_i^\dagger +\int d^2\theta\phi^3
W(Q_i)\, ,
\eeq
where $W(Q_i)$ must have charge 2 to match the charge of $\phi^3$ \footnote{
Indeed the usual $R$-symmetry is a combination of the one here defined
(which commutes with supersymmetry) with the axial symmetry of the superconformal
group. Both symmetries are formally broken by the VEV of the real component
of the compensator.}. Notice that
when $V$ gets a $D$-term 
we get at tree level $m_i^2=-(r_i-2/3)V_D=-q_i V_D$ in
agreement with eq.~(\ref{shiftwz}).

\section{Conclusions}

Theories of supersymmetry breaking induced 
by the conformal anomaly \cite{rs}
present several interesting properties. 
They are simple to realize, very predictive
 and the induced soft masses are independent of flavor physics.
Nevertheless, they have a big drawback: they lead to
 tachyons (in particular, the sleptons).

Here he have reexamined these theories focusing on their
ultraviolet decoupling property.
We have shown that decoupling is lost whenever a 
singlet (a {\it modulus}) remains  light below the high-energy threshold.
These non-decoupling effects have been used to solve the tachyon problem.

In our first model, named anti-GMSB,
the anomaly mediated masses are modified by 
heavy vectorlike fields 
in such a  way that they end up being  positive at low-energy 
(except for the Higgs).
The model is still quite predictive since the spectrum depends
only on the number $N$ and the mass $M$ of the vectorlike fields. Furthermore,
the soft masses are also flavor independent if the mass
of these  fields is below the scale of flavor.
Phenomenologically the model is quite interesting since it presents
a mass spectrum different from other scenarios.
For instance, for $N=4$ the gluino is the lightest gaugino.
The lightest scalar can be the down squark, but it is
more typically the right-handed slepton. Depending on $\mu$ and
$\tan\beta$ the LSP is either the lightest scalar or a combination
of zino and neutral Higgsino. Although
the $\mu$-parameter cannot be predicted, it can be generated
by the simple mechanism explained in section 6. 
Unlike other scenarios, the gravitino is very heavy $\approx 10$ TeV
since it gets its mass at tree-level.

A realistic mass spectrum can also be found in theories with
extra U(1)'s broken at high-energies.
After integrating out the heavy U(1) sector,
non-decoupling remnants can
again avoid the problem with tachyons.
Flavor independence of the soft masses is guaranteed 
in this case if the U(1)-charges are
flavor independent.

In the models considered so far
supersymmetry breaking was originating from
the $F$-term of the gravitational 
multiplet.
In the last section we have generalized this to include a $D$-term
as a new source of supersymmetry breaking.
This requires gauging an $R$-symmetry and consequently enlarging the
MSSM field content.

We want to conclude by emphasizing 
that  the theories considered here make testable predictions
on the sparticle  spectrum.
This encourages us  to further explore their
phenomenology at present and future experiments.

\section*{Acknowledgements}

We acknowledge stimulating discussions with
Gian Giudice, Markus Luty, Massimo Porrati, Alessandro Strumia,
Carlos Wagner and Fabio Zwirner.

\newpage

\def\ijmp#1#2#3{{\it Int. Jour. Mod. Phys. }{\bf #1~}(19#2)~#3}
\def\pl#1#2#3{{\it Phys. Lett. }{\bf B#1~}(19#2)~#3}
\def\zp#1#2#3{{\it Z. Phys. }{\bf C#1~}(19#2)~#3}
\def\prl#1#2#3{{\it Phys. Rev. Lett. }{\bf #1~}(19#2)~#3}
\def\rmp#1#2#3{{\it Rev. Mod. Phys. }{\bf #1~}(19#2)~#3}
\def\prep#1#2#3{{\it Phys. Rep. }{\bf #1~}(19#2)~#3}
\def\pr#1#2#3{{\it Phys. Rev. }{\bf D#1~}(19#2)~#3}
\def\np#1#2#3{{\it Nucl. Phys. }{\bf B#1~}(19#2)~#3}
\def\mpl#1#2#3{{\it Mod. Phys. Lett. }{\bf #1~}(19#2)~#3}
\def\arnps#1#2#3{{\it Annu. Rev. Nucl. Part. Sci. }{\bf #1~}(19#2)~#3}
\def\sjnp#1#2#3{{\it Sov. J. Nucl. Phys. }{\bf #1~}(19#2)~#3}
\def\jetp#1#2#3{{\it JETP Lett. }{\bf #1~}(19#2)~#3}
\def\app#1#2#3{{\it Acta Phys. Polon. }{\bf #1~}(19#2)~#3}
\def\rnc#1#2#3{{\it Riv. Nuovo Cim. }{\bf #1~}(19#2)~#3}
\def\ap#1#2#3{{\it Ann. Phys. }{\bf #1~}(19#2)~#3}
\def\ptp#1#2#3{{\it Prog. Theor. Phys. }{\bf #1~}(19#2)~#3}

\newpage

\setlength{\unitlength}{1cm}
\begin{figure}[htb]
\begin{picture}(12,17.5)
\epsfxsize=19.5cm
\put(-3.,-6.3){\epsfbox{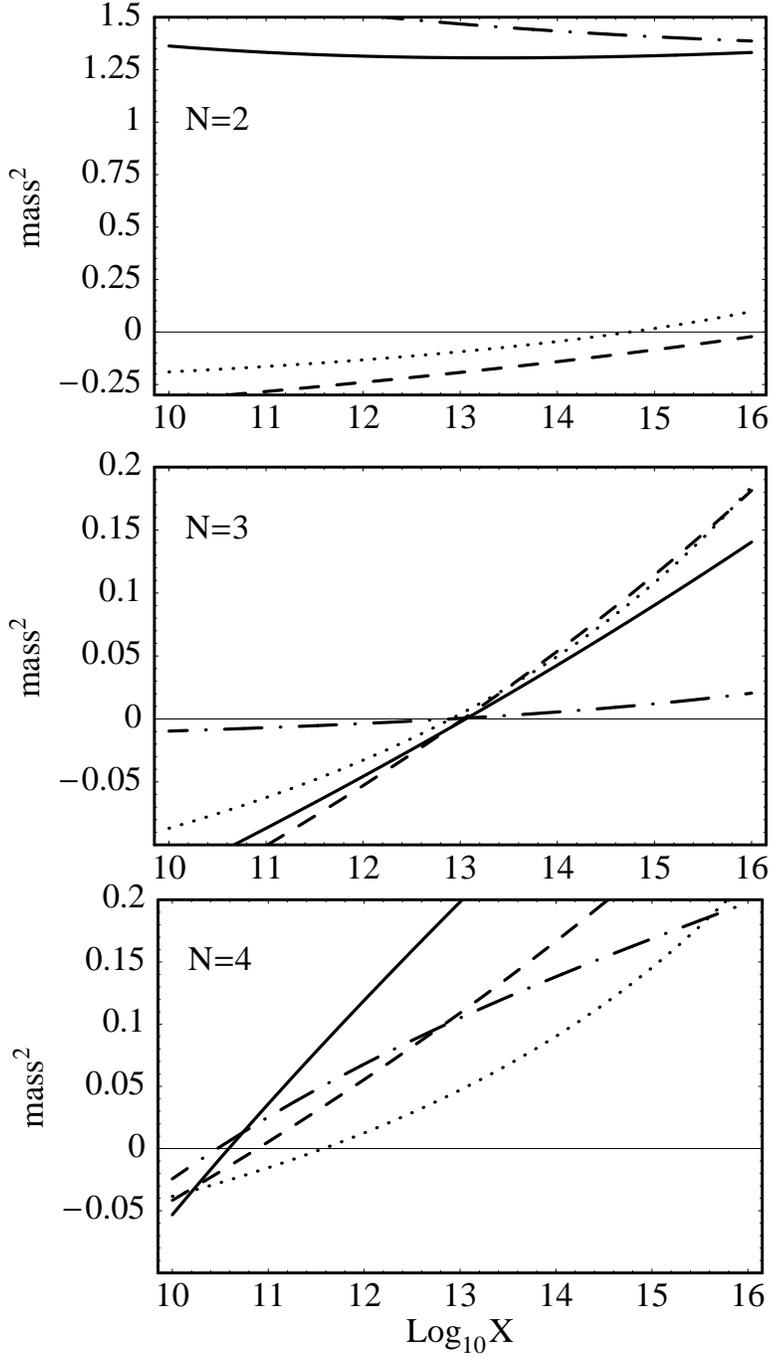}}
\end{picture}
\caption{
Soft squared-masses 
of the left-handed squark (solid line),
 the right-handed down-squark (dashed-dotted line),
the left-handed slepton (dashed line) 
and the right-handed slepton (dotted line) normalized to 
 the SU(2)$_L$-gaugino 
squared-mass as a function of Log$_{10} X$
for $N=2,3,4$  messenger multiplets.
\label{Fig. l}}
\end{figure}


\begin{thebibliography}{99}

\bibitem{gravmed} A.H. Chamseddine, R. Arnowitt, and P. Nath, 
\prl{49}{82}{970}; R. Barbieri, S. Ferrara, and C.A. Savoy, \pl{119}{82}{343};
 L. J. Hall, J. Lykken, and S. Weinberg, \pr{27}{83}{2359}.

\bibitem{flavsymm} M. Dine, R.G. Leigh  and A. Kagan, \pr{48}{93}{4269};
M. Leurer, Y. Nir, and N. Seiberg, \np{420}{94}{468}; A. Pomarol and 
D. Tommasini, \np{466}{96}{3}; R. Barbieri, G. Dvali and L.J. Hall, 
\pl{377}{96}{76}.

\bibitem{DNS} M. Dine, A. Nelson and Y. Shirman, \pr{51}{95}{1362};
M. Dine, A. Nelson, Y. Nir and Y. Shirman, \pr{53}{96}{2658}.

\bibitem{grrev} For a review, see
G.F. Giudice and R. Rattazzi, preprint hep-ph/9801271.

\bibitem{rs} L. Randall and R. Sundrum, MIT-CTP-2788, hep-th/9810155.



\bibitem{glmr} G.F. Giudice, M.A. Luty, H. Murayama and R. Rattazzi,
CERN-TH-98-337, hep-ph/9810442.


\bibitem{noscale} E. Cremmer, S. Ferrara, C. Kounnas, and D.V. Nanopoulos, 
\pl{133}{83}{61};
J. Ellis, A.B. Lahanas, D.V. Nanopoulos, and K. Tamvakis, \pl{134}{84}{429};
J. Ellis, C. Kounnas, and D.V. Nanopoulos, \np{241}{84}{406}.


\bibitem{sugra} E. Cremmer, B. Julia, J. Scherk, S. Ferrara,
 L. Girardello
and P. van Nieuwenhuizen, \np{147}{79}{105}; E. Cremmer, S. Ferrara,
L. Girardello and A. Van Proeyen, \np{212}{83}{413}.

\bibitem{aglr} N. Arkani-Hamed, G.F. Giudice, M. Luty and R. Rattazzi,
\pr{58}{98}{115005}.

\bibitem{dred} W. Siegel, \pl{84}{79}{193}.

\bibitem{nsvz} D.R.T. Jones, \pl{123}{83}{45};
V.A. Novikov, M.A. Shifman, A.I. Vainshtein, and V.I. Zakharov,
\np{229}{83}{381}; {\it ibid.}\/, {\bf B260} (1985) 157;
D.R.T. Jones and L. Mezincescu, \pl{151}{85}{219}.

\bibitem{kl} V. Kaplunovsky and J. Louis, \np{422}{94}{57}.

\bibitem{ahm} N. Arkani-Hamed and H. Murayama, preprint hep-th/9707133.



\bibitem{ibanez} L. Iba\~nez and D. Lust, \np{382}{92}{305};
A. Brignole, L. Iba\~nez and C. Mu\~noz, \np{422}{94}{125}.

\bibitem{zwirner} J.P. Derendinger, S. Ferrara, C. Kounnas and F. Zwirner,
\np{372}{92}{145}, \pl{271}{91}{307}; L. Dixon, V. Kaplunovsky and
J. Louis, \np{355}{91}{649}; G. Lopes Cardoso and B. Ovrut,
\np{369}{92}{351}.


\bibitem{ku} T. Kugo and S. Uehara, \np{222}{83}{125}.


\bibitem{ps} J. Polchinski and L. Susskind, \pr{26}{83}{3661};
H.P. Nilles, M. Srednicki, and D. Wyler, 
\pl{124}{83}{337}; A. Lahanas, \pl{124}{83}{341}.

\bibitem{np}
H.P. Nilles and N. Polonsky, \pl{412}{97}{69}.

\bibitem{gr} G.F. Giudice and R. Rattazzi, \np{511}{98}{25}.

\bibitem{cw} S. Coleman and E. Weinberg, \pr{7}{73}{1888}.


\bibitem{dp} G. Dvali and A. Pomarol, \np{522}{98}{3}.

\bibitem{dgp} G. Dvali, G.F. Giudice, and A. Pomarol, \np{478}{96}{31}.

\bibitem{giumas} G.F. Giudice and A. Masiero, \pl{206}{88}{480}.




\bibitem{gaugeR} A.H. Chamseddine and H. Dreiner, \np{458}{96}{65};
D.J. Casta\~no, D.Z. Freedman and C. Manuel, \np{461}{96}{50}.


\bibitem{fgkv} S. Ferrara, L. Girardello, T. Kugo and A. Van Proeyen,
\np{223}{83}{191}.
\end{thebibliography}
\end{document}